\documentstyle[preprint,aps,prb,floats,epsfig]{revtex}
\tightenlines
\begin{document}
\title{Memory and superposition in a spin glass}

\author{R. Mathieu, P. J\"onsson, D. N. H. Nam\thanks{Also Institute of Materials Science, NCST, Nghiado - Caugiay - Hanoi, Vietnam}, P. Nordblad}
\address{Department of Materials Science, Uppsala University, Box 534, SE-751 21 Uppsala, Sweden}

\date{\today}

\maketitle

\begin{abstract}
Non-equilibrium dynamics in a Ag(Mn) spin glass are investigated by
measurements of the temperature dependence of the remanent
magnetization.  Using specific cooling protocols before recording the
thermo- or isothermal remanent magnetizations on re-heating, it is
found that the measured curves effectively disclose non-equilibrium
spin glass characteristics such as aging and memory phenomena as well
as an extended validity of the superposition principle for the
relaxation.  The usefulness of this ``simple'' dc-method is discussed,
as well as its applicability to other disordered magnetic systems.
\end{abstract}


\section{Introduction}

The non-equilibrium nature of spin glasses and other glassy
systems at low temperature implies that a comprehensive study of the
dynamics usually requires sophisticated dc or ac relaxation
experiments\cite{rev1,rev}.  Dc relaxation is typically used to investigate
the aging properties in spin glasses.  In such experiments, the sample
is first cooled from above the transition temperature $T_g$ to a
constant temperature. After a waiting time $t_w$, a small dc field is
applied (or cut off), and the magnetization is recorded versus time.
The relaxation occurs in a characteristic way, and shows a clear
dependence on $t_w$\cite{aging,rev1,rev}.  Relaxation of the low-frequency ac
susceptibility\cite{jonason,jonsson2,Vincent} is a prime tool
to investigate memory effects in glassy magnets.  When cooling from
above $T_{g}$, a halt at constant temperature $T_{s}$$<$$T_{g}$ is made
during $t_{s}$, allowing the system to relax towards its equilibrium
state at $T_{s}$; both components of the ac-susceptibility then decay
in magnitude.  This equilibrated state becomes frozen in on further
lowering the temperature, and is retrieved on re-heating.  The weak
low frequency ac-field employed in this kind of experiments does not
affect the non-equilibrium processes intrinsic to the sample, but only
works as a non-perturbing probe of the system.\\

In the present paper, an alternative method to investigate
non-equilibrium dynamics is used; ``simple''
remanent magnetization measurements on a Ag(Mn) spin glass
sample.  The magnetization curves are recorded on heating after
specific cooling protocols.  The comparison of the different curves
yields information on dynamic properties such as aging and memory
phenomena.

\section{Memory and superposition}

The sample used in the experiments is a dilute magnetic alloy: the
archetypical three dimensional spin glass Ag(11 at\% Mn).  This
material is chosen as an established reference system to certify the validity
of the dc method presented here.\\
The sample was prepared by melting pure Ag and Mn together at
$T=1000^{\circ}$C in an evacuated atmosphere.  After annealing the
sample at $850^{\circ}$C for 72 h it was quenched to room temperature.
The experiments were performed in a non-commercial low-field SQUID
magnetometer \cite{SQUID}.  The dc magnetic field is generated by a
small superconductive solenoid coil always working in persistent mode
during measurements. The background field was less than 1 mOe.

\begin{figure}
\centerline{\epsfig{figure=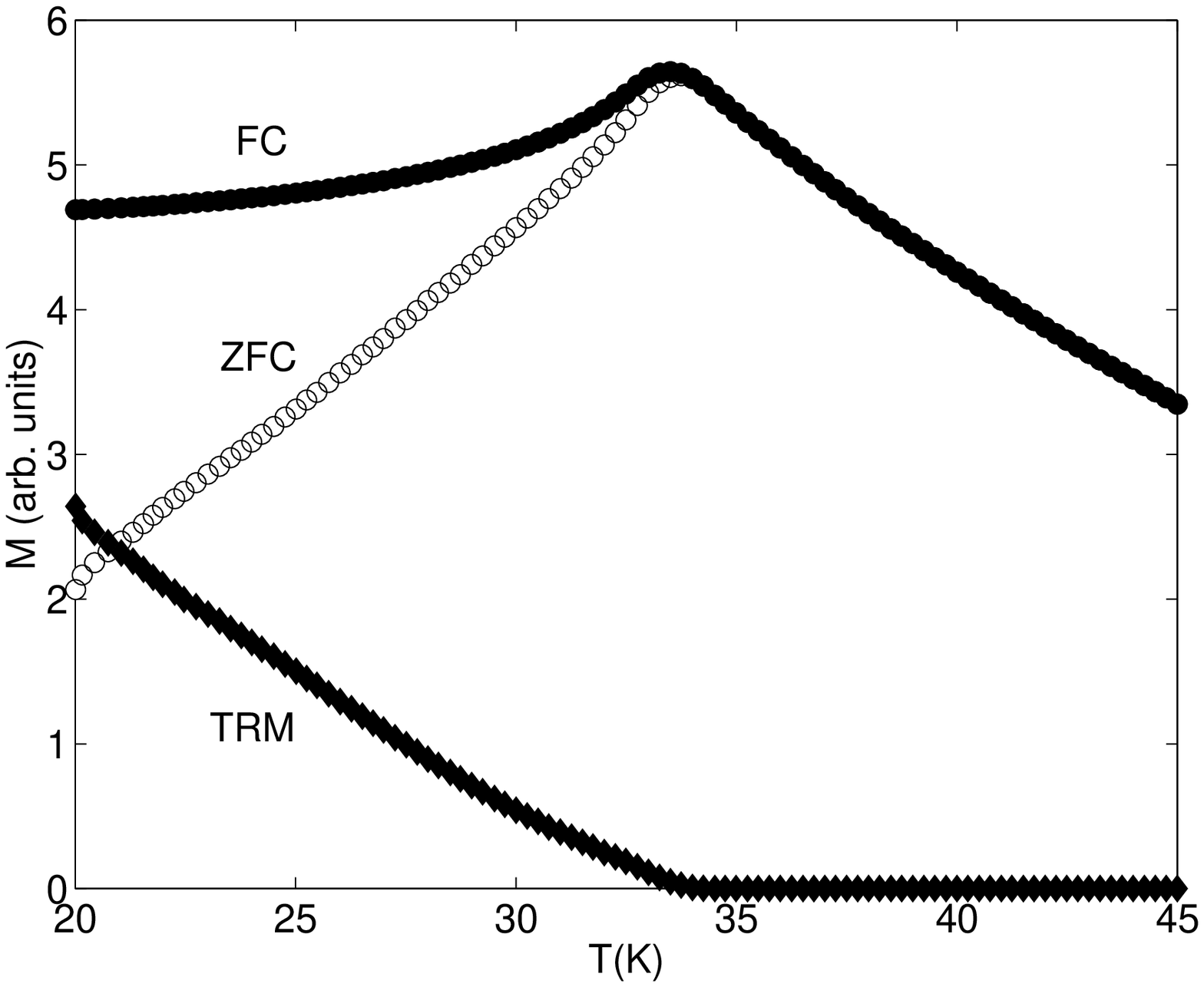,height=6cm,width=8cm}}
\caption{ZFC, FC and TRM magnetizations for $H$=0.1 Oe}
\label{fig1}
\end {figure}

\begin{figure}
\centerline{\epsfig{figure=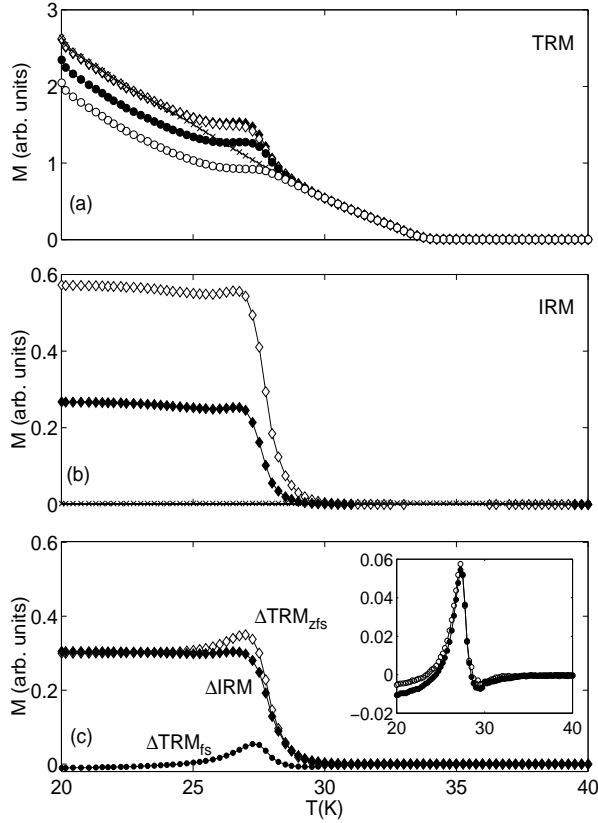,height=11cm,width=8cm}}
\caption{TRM (a) and IRM (b) magnetizations measured after using
different cooling protocols; T$_s$=27K, $H$=0.1 Oe.  (c) displays the
difference between pairs of curves shown in (a) and (b): $\Delta$IRM,
$\Delta$TRM$_{fs}$ and $\Delta$TRM$_{zfs}$.  The inset shows the
difference ($\Delta$TRM$_{zfs}$ - $\Delta$IRM) and $\Delta$TRM$_{fs}$
(which is marked by the same symbol as in the main frame)}
\label{fig2}
\end {figure}

As seen in Fig.~\ref{fig1}, the zero field cooled (ZFC) magnetization
exhibits a characteristic cusp at T=33K, which also roughly defines
the spin glass temperature $T_g$ of our sample. Fig.~\ref{fig2} presents the temperature dependence of (a) the
thermo-remanent (TRM) and (b) the isothermal remanent (IRM) magnetization.
Both were measured on heating in zero field from a low temperature.
The different curves have been recorded after cooling the sample using
the protocol described below:\\

1. The sample is cooled in constant field $H_{o}$=0.1 Oe (TRM) or zero
field (IRM) from a reference temperature above $T_{g}$ to a stop
temperature $T_{s}$ ($T_{s}$ $<$ $T_g$) , where the sample is kept a stop time $t_{s1}$
without changing the field.

2. A field change is made: In the TRM case the field is either cut
to zero (zero-field stop, ZFS) or kept constant (field stop, FS);
for the IRM, $H_{o}$ is applied. The new magnetic field value is
kept during a time $t_{s2}$.

3. The field is then shifted back to its initial value and the
sample is immediately cooled to a lowest temperature $T_{min}$,
where

4. the field is cut to zero (TRM) (or kept at zero (IRM))
and the remanent magnetization $M_{\mathrm{TRM}}(T)$ (denoted
$\mathrm{TRM_{ZFS}}$ and $\mathrm{TRM_{FS}}$ for the case of ZFS
and FS, respectively) or $M_{\mathrm{IRM}}(T)$ (denoted IRM) is
measured on heating the sample at a constant heating rate.\\

\begin{figure}
\centerline{\epsfig{figure=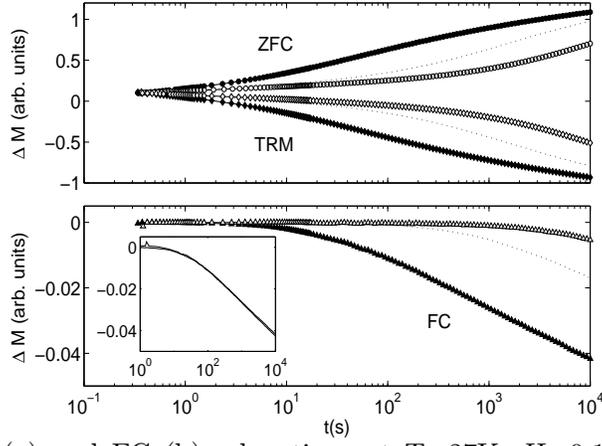,height=6cm,width=8cm}}
\caption{ZFC, TRM (a) and FC (b) relaxations at T=27K, $H$=0.1 Oe for
different waiting times: t$_w$=0s (filled symbols), t$_w$=1000s
(dotted lines) and t$_w$=10$^4$s (open symbols).  The insert shows the
agreement between ZFC+TRM and FC relaxations (t$_w$=0s)}
\label{fig3}
\end {figure}

The different curves in Fig.~\ref{fig2} are measured for two different values of
$t_{s1}$: 0 s and 10000 s, respectively, and in all cases
$t_{s2}=10000$ s; $T_{s}$=27K; $T_{min}$=20 K. The reference
curves, without any field stops, are added. Figs.~\ref{fig2} (a) and (b) show
that the 0 s $\mathrm{TRM_{ZFS}}$
curve lies significantly below the 10000 s curve whereas the 0 s IRM
curve lies well above the 10000 s one.  These curves reflect the paramount
influence that aging has on the magnetic relaxation in spin glasses. 
For comparison, Fig.~\ref{fig3} 
presents parallel dc-relaxation experiments (ZFC, FC and TRM), performed at the
temperature $T_{s}$, using the same small magnetic field, for waiting
times of $t_w$=0 s and $t_w$=10000 s; the results for $t_w$=1000 s are
added to show the continuity of the waiting time dependence. The
observation time, $t$, is defined as the time elapsed after the field
application (cut off).\\
Both $\mathrm{TRM_{FS}}$ curves in Fig.~\ref{fig2} (a) lie
significantly above the reference curve, conveying that a considerable
reinforcement of the spin structure occurs during the stop time
$t_{s1} + t_{s2}$.  The effect even overcomes the downward
relaxation of the FC magnetization occurring during the stop at 27K
(cf. Fig.~\ref{fig3}).  One notices that the difference between
stopping 10000 and 10000+10000=20000 s at $T_{s}$ is significant but
comparably small.\
Provided the experiments are made at low enough field, where there
is a linear response to a field change, one can relate the
results from different relaxation experiments through the
principle of superposition\cite{Lundgren,djurberg}. This implies e.g. that:
\begin{displaymath}
M_{\mathrm{ZFC}}(t_{w},t)=M_{\mathrm{FC}}(0,t_{w}+t)-M_{\mathrm{TRM}}(t_{w},t).
\end{displaymath}
Fig.~\ref{fig2} (c) shows difference plots, $\Delta \mathrm{TRM_{ZFS}}$
and $\Delta$IRM, of the $\mathrm{TRM_{ZFS}}$ and IRM curves shown in
Figs.~\ref{fig2} (a) and (b) giving a direct measure of the frozen in
excess magnetization due to aging.  Also plotted in the figure is a
difference plot of the $\mathrm{TRM_{FS}}$ curves: $\Delta
\mathrm{TRM_{FS}}$.  The inset shows a plot of $\Delta
\mathrm{TRM_{ZFS}}$-$\Delta$IRM and $\Delta \mathrm{TRM_{FS}}$, the
two curves look very similar and suggest an extended validity of the
superposition principle to apply also to the temperature dependence of
the frozen in excess magnetization of a spin glass.  (IRM reflects
$M_{\mathrm{ZFC}}$, $\mathrm{TRM_{ZFS}}$ reflects $M_{\mathrm{TRM}}$,
and $M_{\mathrm{FC}}$ corresponds to $\Delta \mathrm{TRM_{FS}}$.)  The
applicability of the ``ordinary'' superposition principal for the
relaxation of our spin glass is illustrated in Fig.~\ref{fig3}; the
insert shows the sum of the ZFC and TRM relaxations obtained for
$t_w$=0 s and the FC relaxation for the same waiting time.  The curves
are plotted without markers to be able to visually distinguish that
there are two of them; as expected, they are closely equal to each
other.

\begin{figure}
\centerline{\epsfig{figure=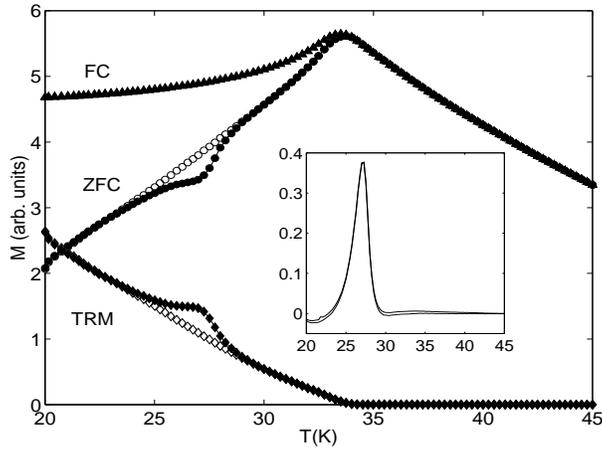,height=6cm,width=8cm}}
\caption{ZFC, FC and TRM magnetizations for $H$=0.1 Oe.  Two curves
(ZFC and TRM) measured after a t=10$^4$s stop at T=27K while cooling
are added; the inset shows the difference with the corresponding
reference}
\label{fig4}
\end {figure}

We have also studied the effect of simple field stops in the ZFC and
FC magnetizations.  Fig.~\ref{fig4} shows the ZFC, FC and TRM
reference curves already presented in Fig.~\ref{fig1}, as well as the
same curves recorded on re-heating after a stop of duration
$t_s$=10$^4$ s in constant field at $T_s$=27K. In the TRM and FC case,
the field is thus kept to its $H$=0.1 Oe value, while for the ZFC, it
remains zero.  The FC curve is only weakly affected by the
stop\cite{jonsson}, whereas the ZFC and TRM curves are
considerably affected\cite{djurberg2}.  The curves reflect the reinforcement of the
spin structure that occurs during the stop at constant temperature and
that this reinforcement sustains when the temperature is recovered on
re-heating.  The insert of Fig.~\ref{fig4} shows the difference
between the FS and reference curves for ZFC, TRM and FC, the excess
magnetization gained due to the FS in TRM + FC corresponds to the
magnetization ``lost'' in the ZFC. The extended validity of the
principal of superposition is further supported by this apparent
agreement.

Looking at the bump and dip in the TRM and ZFC magnetization curves of
Fig.  4 and the derived excess magnetization in the inset, one
observes that the influence of the stop at constant temperature during
cooling is limited to a restricted temperature range around $T_{s}$.
The width of this region may be assigned to the existence of an overlap
between the spin configuration attained at $T_{s}$ and the
corresponding state at a very neighboring temperature ($T+\Delta T$).  The two
concepts that explain the width are then a chaotic nature of the spin
glass equilibrium configuration and an overlap on short length scales
between the equilibrium configurations at $T$ and $T+\Delta T$.  These
concepts have been elaborated earlier \cite{jonason,jonsson2}.  Further
information on these matters can be obtained also by extending the
current dc-method to include two stops when cooling, first at $T_s$ and then at
$T'_s$$<$$T_s$.  The results of this cooling protocol
emphasize the chaotic nature of the equilibrium states at the two
temperatures: the equilibration at $T_s$ has no influence on the
non-equilibrium processes occurring only a few kelvins below, outside
the region of overlap.\\
The actual spin configuration of a spin glass is controlled by its
thermal history.  In this paper we have shown that it is possible to
extract substantial information on complex non-equilibrium phenomena
linked to the evolving spin configuration by simply recording the
temperature dependence of the remanent magnetization after employing
certain cooling protocols. Introducing an excess magnetization in
the spin glass by intermittently applying or removing a magnetic field
during a stop at constant temperature it is possible to study memory
phenomena and the influence of aging on the relaxation of the
magnetisation.  The magnitude of the excess magnetization is governed
by the duration of the field cycle and the wait time at constant
temperature before the magnetic field is changed.  The logarithmic
nature of the relaxation and the aging processes can be explored by
systematically altering the duration of the field cycle and/or the
wait time.  An extended validity of the principal of superposition to
apply to frozen in relaxation processes that are thermally activated
on heating is disclosed by combining results from TRM, ZFC and FC
experiments.

A note on the experimental procedure: the relaxation of spin glasses
extends to eternal times and has a logarithmic character and the
relaxation processes are thermally activated.  These properties
require that the cooling or heating rates in the experiments are
controlled and kept the same in all different measurements to achieve
comparable experimental data.  The heating (and cooling) rate
determines, due to the overlap between equilibrium states at different
temperatures, an effective age of the system and defines an effective
observation time (of order 10 s in our experiments) in the measurement
of the remanent magnetization.

\section{Conclusion}

The present results from the here introduced dc method have been
obtained on an archetypical 3D spin-glass.  Corresponding remanent
magnetization studies can advantageously be performed even using standard SQUID magnetometers.
For example, the slow dynamics of the less conventional La$_{1-x}$Sr$_x$CoO$_3$ glassy system has been successfully
investigated in this way,\cite{lsco005} exposing characteristic memory and superposition features.

\acknowledgments

Financial support from the Swedish Natural Science Research
Council (NFR) is acknowledged. Marie Vennstr\"om is acknowledged for
helping us fabricate the Ag(Mn) sample. D. N. H. Nam is thankful to SAREC/Sida and ISP.

\pagebreak

\end{document}